\documentclass[aps,prl,twocolumn,showpacs,superscriptaddress,a4paper,floatfix]{revtex4-1}

\usepackage{graphicx}
\usepackage{dcolumn}
\usepackage{bm}

\font\scripti=cmmi7
\font\scriptscripti=cmmi5
\def\sib#1{\setbox0 = \hbox{\scripti #1}
  \kern-.02em\copy0\kern-\wd0
  \kern.04em\box0} 
\def\ssib#1{\setbox0 = \hbox{\scriptscripti #1}
  \kern-.02em\copy0\kern-\wd0
  \kern.04em\box0} 
\font\tenib=cmmib10 
\skewchar\tenib='177 \skewchar\tenib='177 \skewchar\tenib='177
\def\pbold#1{\setbox0 = \hbox{$ #1 $}
  \kern-.022em\copy0\kern-\wd0
  \kern.011em\copy0\kern-\wd0
  \kern.011em\copy0\kern-\wd0
  \kern.011em\copy0\kern-\wd0
  \kern.011em\box0} 

\def\up{\uparrow}
\def\dwn{\downarrow}

\def\lesssim{\ \raise.3ex\hbox{$<$}\kern-0.8em\lower.7ex\hbox{$\sim$}\ }
\def\gesim{\ \raise.3ex\hbox{$>$}\kern-0.8em\lower.7ex\hbox{$\sim$}\ }

\begin{document}

\title{Generalized Crossover in Interacting Fermions within the Low-Energy Expansion}

\author{Hiroyuki Tajima}
\affiliation{Quantum Hadron Physics Laboratory, RIKEN Nishina Center (RNC), Wako, Saitama, 351-0198, Japan}
\date{\today}

\begin{abstract}
We generalize the Bardeen-Cooper-Schrieffer-Bose-Einstein-condensation (BCS-BEC) crossover of two-component fermions, which is realized by tuning the $s$-wave scattering length $a$ between the fermions,
to the case of an arbitrary effective range $r_{\rm e}$.
By using the Nozi\`{e}res-Schmitt-Rink (NSR) approach, we show another crossover by changing $r_{\rm e}$
and present several similarities and differences between these two crossovers.
Furthermore, the region ($r_{\rm e}>a/2$) where the effective range expansion breaks down and the Hamiltonian becomes non-Hermitian is found, being consistent with the Wigner's causality bound.
Our results are universal for low-density interacting fermions with low-energy constants $a$ and $r_{\rm e}$ and
are directly relevant to ultracold Fermi atomic gases as well as dilute neutron matter.
\end{abstract}


\maketitle

The BCS-BEC crossover, which is realized by tuning the $s$-wave scattering length $a$ in cold atom experiments~\cite{Giorgini,Bloch,Chin,Regal,Zwierlein,Kinast,Bartenstein}, has been widely accepted as an important concept to understand strongly correlated quantum systems ~\cite{Eagles,Leggett,Nozieres,SadeMelo,Ohashi2,Ohashi3,Chen,Haussmann,Strinati}.
Indeed, this phenomenon has been discussed in various systems such as superconductors ~\cite{Kasahara1,Kasahara2,Yang,Hanaguri,Guidini,Chubukov,Tajima2b,Iskin} and dense quark matter~\cite{Nishida,Abuki2,Abuki1,He}. 
In this regard, thermodynamic properties of strongly interacting ultracold Fermi gases have been experimentally investigated by changing $a$ near the unitarity limit ~\cite{Luo,Horikoshi2,Ku,Navon,Nascimbene,
Horikoshi,Tajima3,Horikoshireview}.
The observed quantities are universal for homogeneous two-component fermions with a dimensionless coupling parameter $1/(k_{\rm F}a)$ where $k_{\rm F}$ is the Fermi momentum.
The equation of state in this atomic system ~\cite{Horikoshi,Tajima3,Horikoshireview,Gezerlis,Forbes,Lacroix,Pieter1} shows an excellent agreement with a variational calculation for dilute pure neutron matter (PNM)~\cite{FP,APR,Gezerlis2} which has a relatively large negative scattering length $a=-18.5$ fm~\cite{AV18} and the coupling parameter $1/(k_{\rm F}a)\simeq -0.04$ at a subnuclear density $\rho\simeq0.08$ fm$^{-3}$.
\par
On the other hand, there are of course various differences between ultracold Fermi gases and pure neutron matter such as non-locality of the interaction.
The most important difference is the magnitude of an effective range $r_{\rm e}$.
While $r_{\rm e}$ in ultracold Fermi gases near a broad Feshbach resonance is negligible, that in neutron matter given by $r_{\rm e}=2.8$ fm~\cite{AV18} largely affects system's properties even around the subnuclear density \cite{Pieter1}. 
On the other hand, a narrow Feshbach resonance in ultracold atoms gives a large and {\it negative} effective range~\cite{Hazlett}.
Since $r_{\rm e}$ is directly related to the phase shift $\delta(\bm{p})$ (where $\bm{p}$ is the momentum), one can expect that the negative (positive) effective range induces a strong (weak) attraction~\cite{Andrenacci,Fregoso,Parish,Ho,Pieter1,Musolino}.
In this sense, a natural question arises: {\it How does the superfluid transition behave if one arbitrarily changes the effective range?}
\par
The purpose of this work is to answer this question 
and show that another crossover of the superfluid phase transition from the BCS pairing to the molecular BEC occurs when changing the effective range $r_{\rm e}$.
We address the superfluid critical temperature $T_{\rm c}$, which is one of the most important topics not only in cold atom physics but also in nuclear physics (for review, see Ref.~\cite{Page}).
For this purpose, we use the Nozi\`{e}res-Schmitt-Rink (NSR) theory~\cite{Nozieres,SadeMelo,Ohashi2,Ohashi3} which is a promising approach to describe the BCS-BEC-crossover physics in the presence of strong pairing fluctuations.
\begin{figure}[t]
\begin{center}
\includegraphics[width=6.5cm]{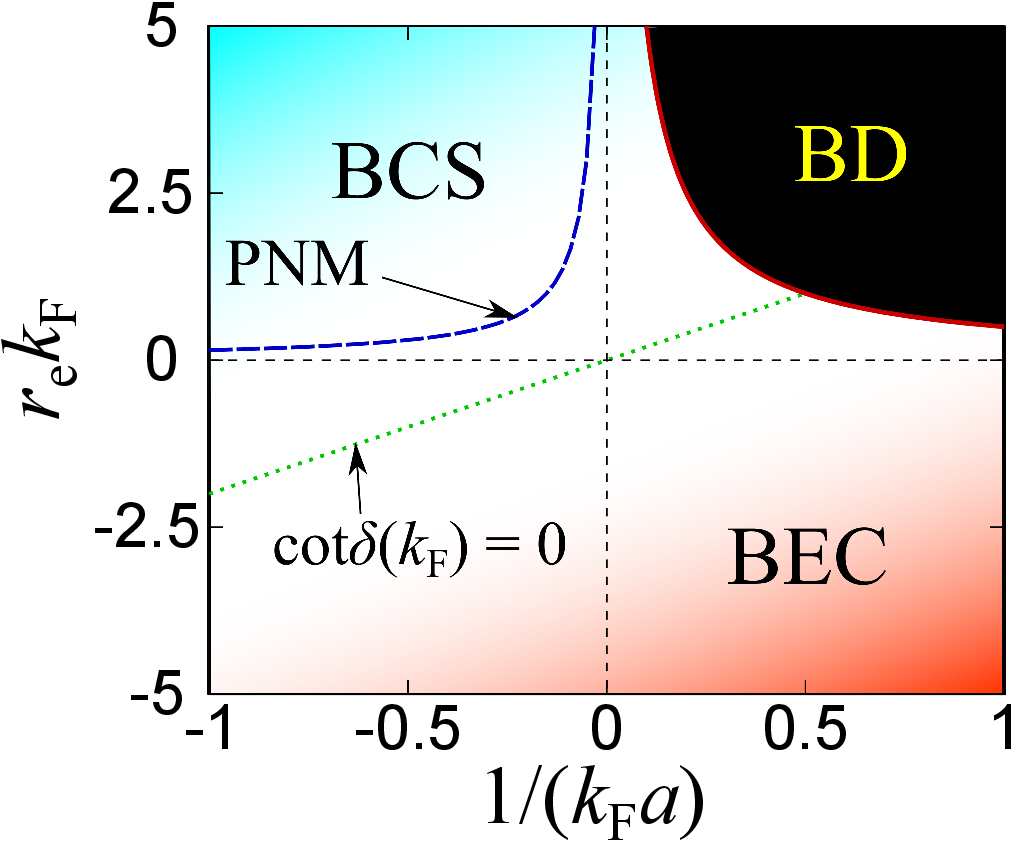}
\end{center}
\caption{(Color online) Phase diagram with respect to the scattering length $a$ and the effective range $r_{\rm e}$ where $k_{\rm F}$ is the Fermi momentum of non-interacting fermions.
The fulfilled area (BD) represents the region ($r_{\rm e}>a/2$) where the effective range expansion breaks down.
The dashed line shows the parameters corresponding to the dilute pure neutron matter (PNM) where $r_{\rm e}/a=-2.8/18.5$.
We also plot ${\rm cot}\delta(k_{\rm F})=0$ indicating an eye-guide of crossover boundary between the BCS and the BEC sides. 
}
\label{fig1}
\end{figure}
Figure \ref{fig1} is a proposed phase diagram of strongly interacting homogeneous spin-$1/2$ fermions with respect to two dimensionless parameter $1/(k_{\rm F}a)$ and $r_{\rm e}k_{\rm F}$.
While the BCS and the BEC sides are qualitatively separated at $1/(k_{\rm F}a)=0$ in the zero-range case, such a crossover boundary is generalized to ${\rm cot}\delta(k_{\rm F})=-1/(k_{\rm F}a)+r_{\rm e}k_{\rm F}/2=0$.
PNM corresponds to the region $r_{\rm e}k_{\rm F}>0$ and $1/(k_{\rm F}a)<0$ where the neutron Fermi momentum $k_{\rm F}$ can vary from the inner crust to the outer core of a neutron star interior. 
In addition, we show that the solution for $T_{\rm c}$ disappears in the large-positive-effective-range region with finite positive scattering length due to the breakdown of the effective range expansion at $r_{\rm e}>a/2$.
Since an independent optical control of the scattering length and the effective range can be achieved in ultracold atomic gases~\cite{Bauer,Wu2,Wu,Semczuk,Jagannathan,Arunkumar},
one can expect that our results can be checked by future experiments.
\par
\begin{figure}[t]
\begin{center}
\includegraphics[width=6.5cm]{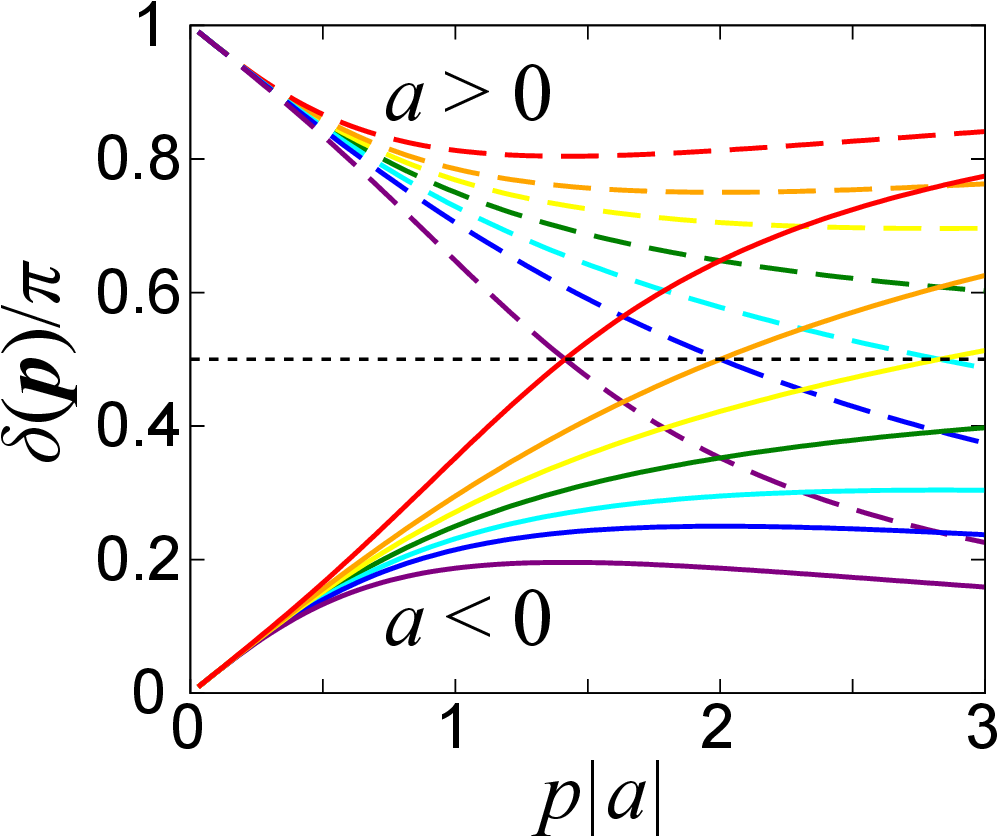}
\end{center}
\caption{(Color online) Phase shifts with negative (solid curves) and positive (dashed curves) scattering lengths $a$. In order from the top of both types of the curves, the effective range is chosen as $r_{\rm e}/a=-1$, $-0.5$, $-0.25$, $0$, $0.25$, $0.5$ and $1$.
The horizontal dotted line ($\delta(\bm{p})=\pi/2$) corresponds to the unitarity limit ($1/a=0$).   
} 
\label{fig2}
\end{figure}

We consider a homogeneous two-component fermion system with low-energy $s$-wave scattering described by a two-channel Hamiltonian~\cite{Holland,Ohashi2,Ohashi3,Liu,Tajima5}
\begin{eqnarray}
\label{eq1}
H&=&\sum_{\bm{p},\sigma}\xi_{\bm{p}}c_{\bm{p},\sigma}^{\dag}c_{\bm{p},\sigma}
+\sum_{\bm{q}}\left(\varepsilon_{\bm{q}}/2+\nu-2\mu\right)b_{\bm{q}}^{\dag}b_{\bm{q}} \cr
&&+\sum_{\bm{p},\bm{q}}\left(g_{\bm{p}}b_{\bm{q}}^{\dag}c_{\bm{p}+\bm{q}/2,\up}c_{-\bm{p}+\bm{q}/2,\dwn}+{\rm H.c.}\right),
\end{eqnarray}
where $\xi_{\bm{p}}=\varepsilon_{\bm{p}}-\mu\equiv\frac{p^2}{2m}-\mu$ is the kinetic energy with the momentum $\bm{p}$ and the fermion mass $m$ measured from the chemical potential $\mu$, $c_{\bm{p},\sigma}$ $(b_{\bm{q}})$ fermionic (bosonic) annihilation operator (where $\sigma=\uparrow,\downarrow$ is the fermionic spin state), $\nu$ the energy of a diatomic boson, and $g_{\bm{p}}=g/\sqrt{1+(p/p_{\rm c})^2}$ the momentum-dependent Feshbach coupling with a cutoff parameter $p_{\rm c}$ (noting that we use $\hbar=k_{\rm B}=1$ and the system volume is taken to be one for simplicity).
In this model, the two-body $T$-matrix is obtained as
\begin{eqnarray}
\label{eqt}
T(\bm{p},\bm{p}',\omega_{+})=\frac{g_{\bm{p}}g_{\bm{p}'}}{\omega_{+}-\nu-\sum_{\bm{k}}\frac{g_{\bm{k}}^2}{\omega_{+}-2\varepsilon_{\bm{k}}}},
\end{eqnarray}
where $\omega_{+}=\omega+i\delta$ and $\delta$ is an infinitesimal small positive number.
Eq. (\ref{eqt}) gives an exact relation between the parameters in Eq. (\ref{eq1}) and the low-energy phase shift
\begin{eqnarray} 
\label{eq2}
p{\rm cot}\delta(\bm{p})
&=& -\frac{4\pi}{m}T^{-1}(\bm{p},\bm{p},2\varepsilon_{\bm{p}}+i\delta) \cr 
&\equiv&-\frac{1}{a}+\frac{1}{2}r_{\rm e}p^2+s p^4,
\end{eqnarray}
where
\begin{eqnarray}
\label{eq3}
a=\left[-\frac{4\pi}{m}\frac{\nu}{g^2}+p_{\rm c}\right]^{-1},
\end{eqnarray}
\begin{eqnarray}
\label{eq4}
r_{\rm e}=-\frac{8\pi}{m^2 g^2}+\frac{2}{p_{\rm c}}\left(1-\frac{1}{p_{\rm c}a}\right),
\end{eqnarray}
and
\begin{eqnarray}
\label{eq5}
s=-\frac{4\pi}{m^2 g^2 p_{\rm c}^2}
\end{eqnarray}
are the scattering length, the effective range, and the shape parameter~\cite{Hamada}, respectively.
In this work, we take $g\rightarrow \infty$ ($p_{\rm c}\rightarrow \infty$) when $r_{\rm e}>0$ $(<0)$
such that $s=0$ since we limit ourselves to the effective range expansion.
We note that higher-order coefficients $O(p^5)$ are exactly zero in this model.
We emphasize that although our choice of the interaction model is specific, 
it would be good starting point to understand non-universal behavior, originating from higher-order coefficients. 
Figure \ref{fig2} shows the phase shift $\delta(\bm{p})$ at various scattering lengths and effective ranges.
While the low-momentum region $p|a|\lesssim 0.5$ of $\delta(\bm{p})$ does not depend on $r_{\rm e}$, the high-momentum region is characterized by $r_{\rm e}$.
Since the positive phase shift indicates how the phase of the wave function is drawn to the short-distance region due to the attraction,
one can interpret the magnitude of $\delta(\bm{p})$ as a strength of the attractive interaction.
From this viewpoint, one can expect that a negative (positive) effective range induces effectively strong (weak) attraction between two fermions.
\par
The condition for obtaining $T_{\rm c}$ is given by the Thouless criterion~\cite{Thouless}
\begin{eqnarray}
\label{eq6}
\frac{mg^2}{4\pi a}+2\mu=-\sum_{\bm{p}}g_{\bm{p}}^2\left[\frac{1-2f\left(\xi_{\bm{p}}\right)}{2\xi_{\bm p}}-\frac{m}{p^2}\right],
\end{eqnarray}
where $f(\xi)=[e^{\xi/T}+1]^{-1}$ is the Fermi distribution function.
In the NSR scheme, the chemical potential $\mu$ at a given number density $N$ is obtained by solving~\cite{Ohashi2,Marcelis}
\begin{eqnarray}
\label{eq7}
N&=&2\sum_{\bm{p}}f\left(\xi_{\bm{p}}\right)+2\sum_{\bm{q}}b\left(\varepsilon_{\bm{q}}/2+\nu-2\mu\right)\cr
&-&T\sum_{\bm{q},i\nu_n}\frac{\partial}{\partial\mu}{\rm ln}\left[1-D_{\bm{q}}(i\nu_n)\Pi_{\bm{q}}(i\nu_n)\right],
\end{eqnarray}
where $D_{\bm{q}}(i\nu_n)=\left[i\nu_n-\varepsilon_{\bm q}/2-\nu+2\mu \right]^{-1}$
is the thermal Green's function of a bare molecular boson ($i\nu_n$ is the bosonic Matsubara frequency) and 
\begin{eqnarray}
\label{eq8}
\Pi_{\bm{q}}(i\nu_n)=\sum_{\bm{p}}g_{\bm{p}}^2\frac{1-f\left(\xi_{\bm{p}+\bm{q}/2}\right)-f\left(\xi_{-\bm{p}+\bm{q}/2}\right)}{i\nu_n-\xi_{\bm{p}+\bm{q}/2}-\xi_{-\bm{p}+\bm{q}/2}}
\end{eqnarray}
is the pair-correlation function.
\begin{figure}[t]
\begin{center}
\includegraphics[width=6.5cm]{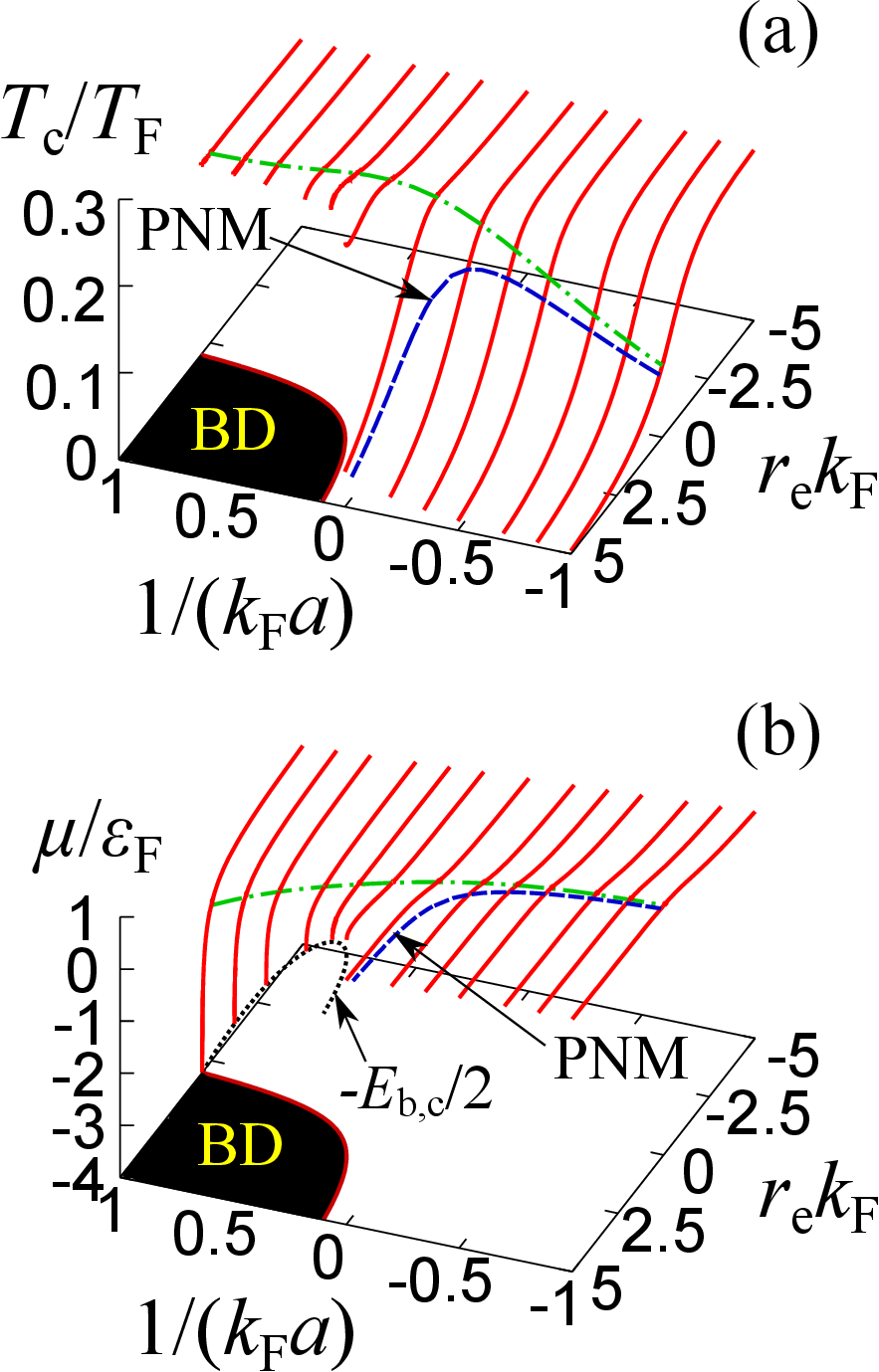}
\end{center}
\caption{(Color online) (a) The superfluid phase transition temperature $T_{\rm c}$ and (b) the chemical potential $\mu$ at $T=T_{\rm c}$ as functions of the inverse scattering length $1/(k_{\rm F}a)$ and the effective range $r_{\rm e}k_{\rm F}$, where $\varepsilon_{\rm F}$ is the Fermi energy of non-interacting fermions.
In each figure,
the dashed and dash-dotted curves show results at PNM parameters and $r_{\rm e}=0$, respectively.
The dotted line in the panel (b) shows the half of the critical binding energy $-E_{\rm b.c}/2=-2/(ma^2)$ at $r_{\rm e}=a/2$. The effective range expansion breaks down in the region beyond $r_{\rm e}=a/2$ (BD).
} 
\label{fig3}
\end{figure}
The third term of Eq. (\ref{eq7}) is the many-body correction associated with strong pairing fluctuations. 
We note that $b(\varepsilon)=\left[e^{\varepsilon/T}-1\right]^{-1}$ is the Bose distribution function.
\par
\label{sec3} 
Figure \ref{fig3} shows the calculated superfluid critical temperature $T_{\rm c}$ and chemical potential $\mu$ at $T=T_{\rm c}$ in the parameter plane of $1/(k_{\rm F}a)$ and $r_{\rm e}k_{\rm F}$.
One can find that the effective-range dependence of $T_{\rm c}$ at fixed $1/(k_{\rm F}a)$ $(\leq 0)$ is similar to the ordinary zero-range results shown as the dash-dotted lines.
$T_{\rm c}$ increases with increasing the magnitude of the negative effective range even at the negative scattering length since the effective interaction is enhanced~\cite{Ho,Tajima5,Marcelis}.
In the weak-coupling regime where $1/(k_{\rm F}a)<0$ and $r_{\rm e}k_{\rm e} >0$, the critical temperature is qualitatively explained by $T_{\rm c}\sim T_{\rm F}{\rm exp}\left[-\frac{\pi}{2}\left(-\frac{1}{k_{\rm F}a}+\frac{1}{2}r_{\rm e}k_{\rm F}\right)\right]$ \cite{Tajima5,Marcelis}, whereas the ordinary BCS prediction with the zero-range potential is given by $T_{\rm c}^{\rm BCS}\simeq 0.614T_{\rm F}{\rm exp}\left(\frac{\pi}{2k_{\rm F}a}\right)$.
In the large-negative-effective-range limit with any finite scattering lengths, $T_{\rm c}$ and $\mu$ approach $T_{\rm LNR}=0.204T_{\rm F}$ and $0$, respectively \cite{Tajima5,Floerchinger}, where the system consists of the BEC of diatomic molecules and a small number of thermal excited fermions given by $N_0=2\sum_{\bm{p}}f(\varepsilon_{\bm{p}})$ which causes the quantitative difference between $T_{\rm LNR}$ and the ordinary BEC limit $T_{\rm BEC}=0.218T_{\rm F}$~\cite{Nozieres,SadeMelo,Chen,Ohashi2,Ohashi3,Haussmann,Strinati}.
On the other hand, $T_{\rm c}$ gradually decreases and $\mu$ approaches $\varepsilon_{\rm F}$ with increasing $r_{\rm e}$ due to the reduction of the cutoff as \cite{Pieter1}
\begin{eqnarray}
\label{eq9}
p_{\rm c}=\frac{1+\sqrt{1-2r_{\rm e}/a}}{r_{\rm e}} \quad (r_{\rm e}>0).
\end{eqnarray}
In this regard, the diagonal effective interaction $U_{\rm eff}(\bm{p},\bm{p})=-g_{\bm{p}}^2/(\nu-2\mu)$ at $\bm{p}=k_{\rm F}$ becomes weaker, resulting in the decrease of $T_{\rm c}$ \cite{Pieter1,Andrenacci,Ramanan1,TajimaANM}.
The pure neutron matter (PNM) corresponding to the dashed curve in Fig. \ref{fig3} is also located in this region, with $a=-18.5$ fm and $r_{\rm e}=2.8$ fm \cite{AV18}.
PNM gradually approaches the unitarity limit ($1/a=0$) with increasing the neutron density $\rho_n=k_{\rm F}^3/(3\pi^2)$ and finally goes to the large-positive-effective-range region.
\par
In contrast, in the region where $a>0$ and $r_{\rm e}>0$,
$r_{\rm e}$ induces the enlargement of the two-body binding energy $E_{\rm b}$ and
$\mu$ approaches the half of $-E_{\rm b}$ given by
\begin{eqnarray}
\label{eq10}
E_{\rm b}&=&\frac{1}{ma^2}\frac{1}{\left[1-1/(ap_{\rm c})\right]^2}\cr
&=&\frac{1}{ma^2}\left[\frac{1+\sqrt{1-2r_{\rm e}/a}}{1-{r_{\rm e}/a}+\sqrt{1-2r_{\rm e}/a}}\right]^2.
\end{eqnarray}
One can find that the real solution of $E_{\rm b}$ disappears if $r_{\rm e}$ excesses $a/2$ since $p_{\rm c}$ given by Eq. (\ref{eq9}) becomes complex.
This result indicates that the effective range expansion breaks down to describe such a deep bound state within the two low-energy parameters.
In such a situation, the Hamiltonian given by Eq. (\ref{eq1}) becomes non-Hermitian
and we cannot obtain a thermodynamic equilibrium state in our homogeneous model. 
We also note that 
while such a singularity was overlooked in the previous mean-field study at $T=0$~\cite{Musolino},
interestingly a cluster formation has been predicted in a similar region where $r_{\rm e}>0.46a$ in trapped Fermi gases~\cite{Yin}.
Indeed, $\mu$ approaches the half of a critical binding energy $-E_{\rm b,c}/2=-2/(ma^2)$ at this boundary $r_{\rm e}=a/2$ ($p_{\rm c}=r_{\rm e}^{-1}$) in the strong-coupling side.
\par
The presence of the upper bound for $r_{\rm e}$ is consistent with the so-called Wigner's causality bound~\cite{Wigner}, which is given by $r_{\rm e}\leq 2r-\frac{2r^2}{a}+\frac{2r^3}{3a^2}$ where 
$r$ is a number being larger than the interaction range $R$~\cite{Hammer1,Hammer2}. 
While we cannot define $R$ in our model,
the causality bound is valid if the interaction potential decreases quickly at a large distance~\cite{Hammer3}.
In the study for two-neutron halo nuclei~\cite{Hammer3},
the Wigner's bound was obtained as the region where the renormalized coupling constants become complex.
Indeed, the obtained bound in this letter can be rewritten as $r_{\rm e}\leq p_{\rm c}^{-1}$,
where $p_{\rm c}^{-1}$ is associated with $R$.
Considering the short-range limit ($p_{\rm c}\rightarrow \infty$),
we can obtain $r_{\rm e}\leq 0$ which is consistent with the previous work~\cite{Hammer1,Hammer2,Hammer3}.  
Moreover, if we set $r=(2-2^{\frac{1}{3}})p_{\rm c}^{-1}$, the causality bound reproduces $r_{\rm e}\leq a/2$.
We note that the upper bound for $r_{\rm e}$ changes in the case with different cutoff functions in $g_{\bm{p}}$.
For instance, the Yamaguchi potential~\cite{Yamaguchi} corresponding to $g_{\bm{p}}^{\rm Y}=g/[1+(p/p_{\rm c})^2]$ requires $r_{\rm e}\leq \frac{9}{16}a$~\cite{TajimaANM} to obtain physical solutions.
Although in this letter we consider the specific model which is characterized by just two parameters $a$ and $r_{\rm e}$, the higher-order coefficients are generally non-zero due to the short-range part of the interaction. 
Actually, $g_{\bm{p}}^{\rm Y}$ also involves non-zero higher-order coefficients beyond the effective-range expansion.
This indicates that the upper bound for $r_{\rm e}$ depends on the non-universal short-range parts. 
We note that although the spin-triplet neutron-proton interaction in dilute nuclear matter has a positive scattering length [$a_{\rm t}=5.42$ fm] and an effective range [$r_{\rm e,t}=1.76$ fm] \cite{AV18}, it does not satisfy $r_{\rm e,t}>a_{\rm t}/2=2.71$ fm and the deuteron (neutron-proton pair) binding energy $E_{\rm d}=2.2$ MeV is consistent with Eq.~(\ref{eq10}) \cite{Yamaguchi,Naidon}. 

\par
\begin{figure}[t]
\begin{center}
\includegraphics[width=6.5cm]{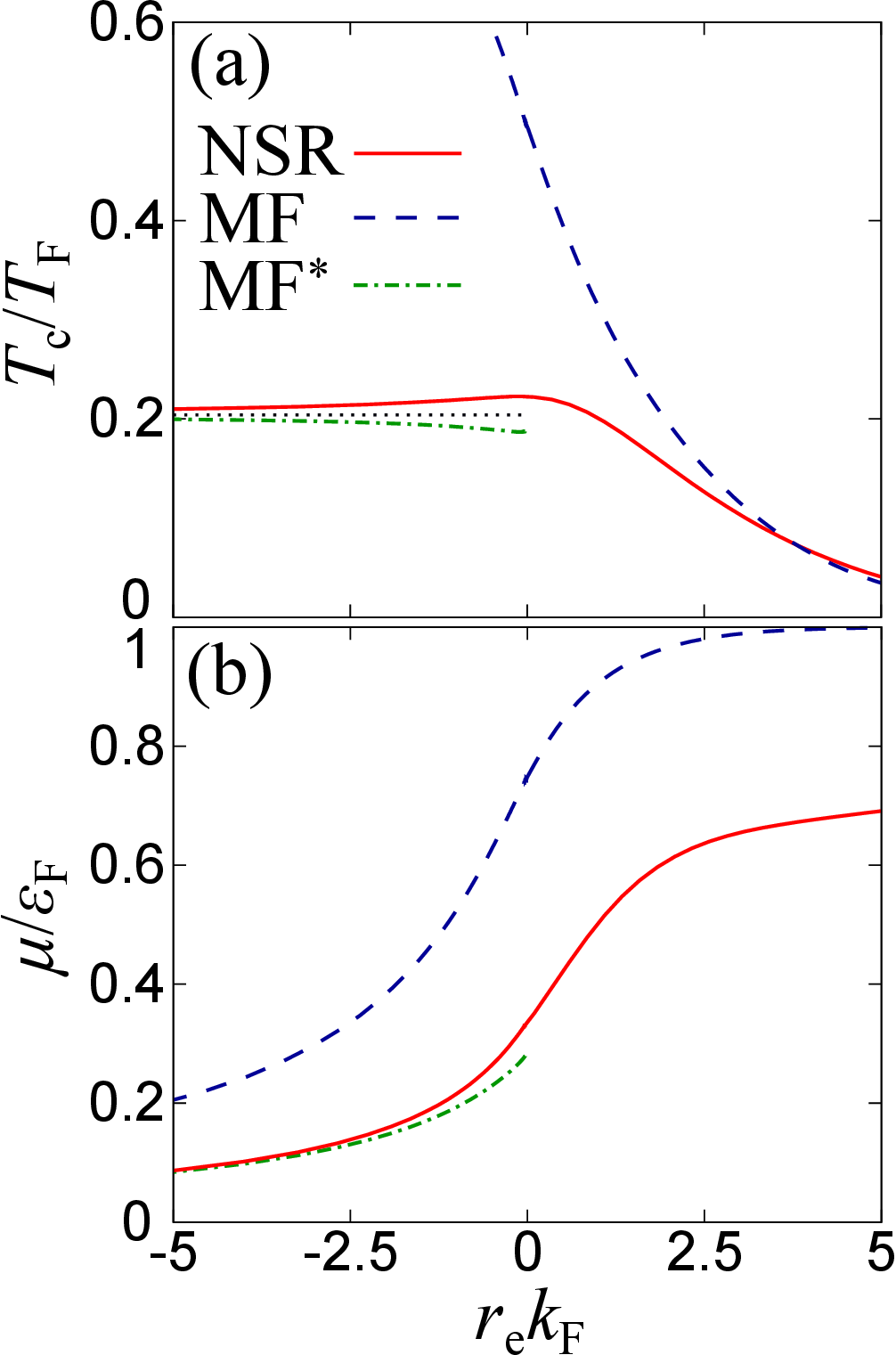}
\end{center}
\caption{(Color online) (a) The superfluid phase transition temperature $T_{\rm c}$ and (b) the chemical potential $\mu$ at $T=T_{\rm c}$ as a function of $r_{\rm e}k_{\rm F}$ at $1/a=0$,
within the NSR and MF approaches.
MF$^*$ is the mean-field results with the ultraviolet renormalization of $\nu$.
The dotted line in the panel (a) shows the large-negative-effective-range limit $T_{\rm LNR}=0.204T_{\rm F}$ \cite{Tajima5,Floerchinger}. 
}
\label{fig4}
\end{figure}
To see the effects of pairing fluctuations on $T_{\rm c}$ and $\mu$,
we compare the results of the NSR approach to the mean-field calculation (MF) at a resonant interaction ($1/a=0$) as shown in Fig. \ref{fig4}.
We note that the mean-field results are obtained by solving Eq. (\ref{eq7}) without the third term associated with strong pairing fluctuations.
As is the ordinary case with a tunable scattering length at $r_{\rm e}=0$,
$T_{\rm c}$ increases and $\mu$ decreases with decreasing $r_{\rm e}k_{\rm F}$
in the mean-field approximation.
In this sense, $\mu$ is qualitatively explained by the mean-field level in the entire crossover region but $T_{\rm c}$ is largely overestimated in the negative-effective-range region.
While the region between the NSR and MF critical temperatures can be regarded as a preformed-pair region in the case of the contact interaction \cite{Tsuchiya,Gaebler,Perali2011},
the depletions of $T_{\rm c}$ and $\mu$ in the negative-effective-range regime are also related to the ultraviolet renormalization of $\nu$. 
Indeed, the mean-field calculation (MF$^*$) with replacing $\nu$ with $\nu_{\rm r}=\nu-g^2\sum_{\bm{p}}\frac{m}{p^2}$ (where $p_{\rm c}\rightarrow \infty$)
shows good agreement with the NSR results, where the number of non-condensed bosons $2\sum_{\bm{q}}b(\varepsilon_{\bm{q}}/2+\nu_{\rm r}-2\mu)$ becomes finite.
Such a simple approximation becomes exact in the large-negative-effective-range limit \cite{Schonenberg}
and this fact is in a sharp contrast with the strong-coupling BEC side realized at small positive scattering length and zero effective range.
We note that this is a direct consequence of the difference between BEC of {\it closed-channel} molecules in the two-channel model at $r_{\rm e}k_{\rm F}\lesssim -2$ and that of {\it tightly bound} molecules  at $1/(k_{\rm F}a)\gesim 1$ with $r_{\rm e}=0$ \cite{Ohashi3}. 
\par
To summarize, we have addressed the generalized crossover within the low-energy expansion involving two interaction parameters, that is, the scattering length $a$ and the effective range $r_{\rm e}$.
By using the Nozi\`{e}res-Schimitt-Rink (NSR) approach to incorporate effects of strong pairing fluctuations, we predict the superfluid phase transition temperature $T_{\rm c}$ at arbitrary dimensionless parameters $1/(k_{\rm F}a)$ and $r_{\rm e}k_{\rm F}$.
The crossover of $T_{\rm c}$ from BCS superfluid to molecular BEC can be achieved by changing $r_{\rm e}$ from the positive value to negative one at $1/a\leq 0$.
In this regime, the positive effective range reduces the effective interaction at the Fermi momentum due to the momentum cutoff $p_{\rm c}\simeq r_{\rm e}^{-1}$.
In the large-negative-effective-range limit at any finite scattering lengths, $T_{\rm c}$ approaches $T_{\rm LNR}=0.204T_{\rm F}$ which is equivalent to the BEC temperature of diatomic molecules in the presence of thermal excited fermions.
In addition, the chemical potential $\mu$ at $T=T_{\rm c}$ shifts from the Fermi energy $\varepsilon_{\rm F}$ to zero by decreasing $r_{\rm e}$.
On the other hand, at $r_{\rm e}>a/2$ with the positive $a$, the effective range expansion breaks down and the physical bound state vanishes, resulting the disappearance of $T_{\rm c}$ and a non-Hermiticity of the model Hamiltonian
in this regime.
\par 
In this work, we have not considered the effects of the Hartree shift which is of importance in the positive-effective-range region \cite{Pieter1}. In particular, it is reported that the Hartree shift with a large positive effective range makes the system thermodynamically unstable at $1/a=0$ \cite{Schonenberg}.
Furthermore, the Gor'kov-Melik-Barkhudarov (GMB) corrections \cite{GMB} on $T_{\rm c}$ associated with particle-hole fluctuations \cite{Floerchinger,Yu,Ruan,Pisani,Pisani2,Ramanan2} play a significant role especially in the weak-coupling region.
These are left as future problems.
\acknowledgements
The author thanks P. Naidon and T. Hatsuda for reading the manuscript and giving useful comments, and S. Endo, Y. Ohashi, and Y. Kondo for stimulating discussions.
This work was supported by a Grant-in-Aid for JSPS fellows (No.17J03975)
and RIKEN iTHEMS program.
\par

\end{document}